%% file: wileySubmissionArxiv.tex
\documentclass{article}

\usepackage{cite}
\usepackage{amsfonts}
\usepackage{bbold}
\usepackage{moreverb}
\usepackage{graphicx}  
\usepackage{mathrsfs}
\usepackage{amsmath}
\usepackage{amssymb}
\usepackage{verbatim}
\usepackage{multirow}

\usepackage{verbatim}

\newcommand{\cS}{\mathcal{S}} 
\newcommand{\cJ}{\mathcal{J}}

\newcommand{\bs}{\mathbf{s}} 
\newcommand{\bz}{\mathbf{z}} 
\newcommand{\iz}{\mathit{z}} 
\newcommand{\by}{\mathbf{y}} 
\newcommand{\iy}{\mathit{y}} 
\newcommand{\bv}{\mathbf{v}} 
\newcommand{\VAR}{\operatorname{Var}} 
\newcommand{\COV}{\operatorname{Cov}} 

\begin{document}

\title{Adapting conditional simulation using circulant embedding for irregularly spaced spatial data}
\date{}
\maketitle
\vspace{-1in}
\begin{center}
{{M. Bailey}, 
{S. Bandyopadhyay}, and
{D. Nychka} \\
Department of Applied Mathematics and Statics, Colorado School of Mines 
}
\end{center}
\abstract{
Computing an ensemble of random fields using conditional simulation is an ideal method for retrieving accurate estimates of a field conditioned on available data and for quantifying the uncertainty of these realizations. Methods for generating random realizations, however, are computationally demanding, especially when the estimates are conditioned on numerous observed data and for large domains. In this article, a \textit{new}, \textit{approximate} conditional simulation approach is applied that builds on  \textit{circulant embedding} (CE), a fast method for simulating stationary Gaussian processes. The standard CE is restricted to simulating stationary Gaussian processes (possibly anisotropic) on regularly spaced grids. In this work we explore two possible algorithms, namely local Kriging and nearest neighbor Kriging, that extend CE for irregularly spaced data points. We establish the accuracy of these methods to be suitable for practical inference and the speedup in computation allows for generating conditional fields close to an interactive time frame.  The methods are  motivated by the U.S. Geological Survey's software \textit{ShakeMap}, which provides near real-time maps of shaking intensity after the occurrence of a significant earthquake.  An example for the 2019  event in Ridgecrest, California is used to illustrate our method.  }



\input{wileySubmissionBody.tex}

\newpage
\include{References/References}

\end{document}

%% file: wileySubmissionBody.tex
\section{Introduction}\label{sec1}
An important contribution of spatial statistics is not only being able to make predictions at unobserved locations but to also attach a measure of uncertainty to these predictions. 
 Conditional simulation is a Monte Carlo based method that is ideal for quantifying  prediction uncertainty at new locations especially when the predictions comprise a large, fine grid of points representing a surface.  Besides providing Monte Carlo estimates of prediction standard errors it can also be applied to  nonlinear features in a spatial prediction such as the uncertainty in contour lines and level sets.  A major computational bottleneck in conditional simulation, however, is the initial simulation of the {\it unconditional} spatial process at a large set of observed and unobserved locations.  Using exact methods limits the analysis on a laptop and in R  to a combined number of locations on the order of tens of thousands. This constraint is easily exceeded for larger spatial data sets and prediction to modest sized spatial grids.   Here we exploit  the circulant embedding (CE) method \cite{wood1994simulation, chan1999} because of its efficiency and make some additions for irregular locations.   The focus of our work is on geophysical problems where the conditional simulation is typically required on a fine grid and where computing resources are modest.  
 
 The value of CE lies in the fact that it is both  a  \textit{fast} and \textit{exact} method for simulating stationary Gaussian random fields on a regular grid. For example, on an $m\times m$ lattice, CE uses $40m^2  \operatorname{log}_22m$ floating point operations  (FLOPS) compared to $6m^5$ FLOPS using a Cholesky decomposition. This reduced operation count translates to significant speedup in simulation time and reduction in memory. If observation locations are a subset of the grid then generating realizations from the conditional distribution is a straight forward computation  by pairing CE unconditional simulation with efficient implementation of Kriging predictions.
  However,  a major restriction of CE is that it only simulates the process on a regular grid of locations. For data sets where the observation locations are irregular and can not be registered to a prediction grid  it is difficult to 
 simulate the process simultaneously at locations both on a grid and at  observation locations. This follows because the efficiency of CE cannot be exploited in this case. We study two simple local algorithms that use CE and extend conditional simulation to irregular, off-grid locations. These extensions produce accurate simulations of the prediction error and take a small fraction of computational time relative to other parts of the conditional simulation algorithm. Although at its core these is are simple modifications ,  to our knowledge  these algorithms have not been identified in previous work and  implemented in open source data analysis environments such as R.
  
 This work is motivated by a practical problem encountered by the United States Geological Survey (USGS) in using earthquake measurements. 
Ground motion sensors, used to measure earthquakes are typically irregularly placed in a spatial domain. Based on these data, estimates of median ground motion are produced by  the USGS software package, {\it ShakeMap} (\cite{worden2020})  in conjunction with observed seismic data \cite{verros2017computing}. Generally, ShakeMap based estimates are then used as input fields in  additional models that quantify fatalities and economic losses from significant seismic events. The statistical problem  is to extrapolate ShakeMap estimates to a fine spatial field to address losses over a complete spatial region and to quantify the uncertainty in these estimates. In particular, several hundred conditional simulations of the field are required to compute a Monte Carlo estimate of the prediction standard error. 

 Throughout this work the two new methods will be   referred to as {\it local Kriging} and {\it nearest neighbor Kriging} and we evaluate them by the accuracy in approximating the exact  prediction variances over a range of spatial data models. There are several strategies in the literature related to this problem.  To work with irregular  grids, a block circulant embedding method has been proposed \cite{BCEM}. This method focuses specifically on data that has a block circulant structure, however, and  is less applicable to the ShakeMap software. A second approach currently used in the R package \texttt{fields} and implemented as the function {\tt sim.mKrig.approx}. Here linear interpolation uses the four nearest grid points to simulate an observation \cite{furrer2009package}. While this strategy is very fast, it does not preserve the covariance structure of the spatial data, does not adjust for interpolation error, and its accuracy is untested.

An elegant solution to simulating unconditional and conditional Gaussian spatial fields is to use algorithms based on sequentially conditioning on subsets of the locations, termed {\it Vecchia}  approximations  (VA) (\cite{katzfuss2020vecchia}). This is a comprehensive and  efficient framework for approximating Gaussian process computations. Although  an alternative for simulation, coding a VA is more complex and it is fundamentally a sequential algorithm that can be more difficult to implement in parallel. Under some circumstances our approach fits into the Vecchia framework. For our local Kriging method if the neighborhoods of grid points do not overlap then we are following a Vecchia algorithm where the final conditioning set is the entire field on the regular grid generated by CE. When neighborhoods overlap our method departs from the standard Vecchia setup and we study the impact of this difference.  In either case our approach benefits from being able to simulate large and exact unconditional fields using CE that moderate the effect of computational shortcuts in simulating at the off grid observation locations.  

Another difference with our work compared to that of using a VA is our focus on conditional simulation and the Monte Carlo approximation of the prediction standard error.   We study how well our methods work in conditional inference for spatial problems and observation densities similar to the Shakemap application and we believe this is an important metric for judging our algorithms.  As a practical example we consider  sensor observations from an earthquake near Ridgecrest, CA in 2019 and work with more than 600 spatial locations and a prediction grid of more than 250K points. The goal is to handle this size problem in an interactive data analysis environment, such as R, and on a standard laptop computer and generate accurate prediction intervals.

In the next section we detail the relevant computational algorithms. This is followed  in Section 3 with a derivation of the misspecification of the prediction standard error under our proposed algorithms. Also numerical results are reported for the approximation error and timing results for a spatial context relevant to USGS data. Section 4 applies the method to the Ridgecrest, CA seismic event and Section 5 closes with some discussion of future directions.

\section{Extending Circulant Embedding Simulation}

In this work  we assume a stationary,  mean-zero spatial process $\mathit{y}(\cdot)$ with finite variance $\sigma^2$ and associated stationary covariance function, $C(\cdot)$. For example, $\mathit{y}$ may represent a particular shaking intensity measurement such as log peak ground acceleration.  We assume the spatial process is observed with error at $\mathit{n}$ spatial locations $\{ \bs_i, i=1,\ldots,n \}$ and can be represented as  the standard additive geostatistical model:
\begin{equation*}
    z_i =\mu(\bs_i)+y(\bs_i)+\varepsilon_i,\ \mbox{for} \ i = 1, \ldots n.\label{standardgeostat}
\end{equation*}

\noindent  In the above equation, $\mathit{z}_i$ is the observation at location $\bs_i$, $\mu(\cdot)$ is the mean function, and
$\varepsilon_{i}$'s  are mean zero, Gaussian random variables with
variance $\phi_i^2$, often referred to as the \textit{nugget
effect}. It is assumed the $\varepsilon$ is uncorrelated with
$\mathit{y}$. The mean function typically takes the form of a linear model, 
$\mu(\bs_i)=\boldsymbol{x}(\bs_i)^T\boldsymbol{\beta}$, where
$\boldsymbol{x}(\bs)=[x_1(\bs), \ldots, x_n(\bs)]^T$ and the
parameters $\boldsymbol{\beta}$ are estimated using
generalized least-squares (see \cite{cressie1993}). To streamline the
presentation, it  is assumed that $\mu(\bs_i)$ is zero.
The extension to include a linear mean function is straight forward
and is mentioned in the discussion. Finally, we note that here "measurement error" is a generic component that can include microscale variation in the spatial process or essentially any component of the observation that is not predictable from other spatial locations. 

Given $\bz$ we are interested in predicting $\mathit{y(\bs)}$ on the regularly spaced prediction grid $\mathcal{S}^g=\{\bs_1^g, \bs_2^g, \ldots, \bs_{M}^g\}$. Specifically in 2 dimensions, for an $m_1 \times m_2$ grid  $M= m_1 m_2$. 
In this work, it will be convenient to use  notation $\mathit{y}(A)=\{\mathit{y}(\pmb{\ell}): \pmb{\ell}\in A\}$ to denote the {\it vector} of $\mathit{y}$ variables over the set $A=\{\pmb{\ell}_1, \ldots,\pmb{\ell}_k\}$ and the order in this vector is fixed.  Thus, the computational goal is 
to simulate values for $y(\cS^g)$ conditioned on a spatial process observed with error,  $\bz=[\iz_{1},\ldots,\iz_{n}]^{T}$. 
The conditional simulation  algorithm  has the following steps
 assuming a Gaussian process for $\mathit{y}$ and normally distributed error: 
    %
    \begin{enumerate}
        \item Compute the spatial prediction of the process at the prediction grid  based on the actual spatial data. Denote this by $\widehat{\by}$.
        \item Simulate a spatial process $\iy$ at the union of locations $S=\cS^g \cup \cS$. 
        \item Form the synthetic  data $\iz_i^* =\mathit{y}(\bs_i)+\varepsilon_i^* $, for  $\bs_i\in \cS$ and $\varepsilon_i^*$ are independent $N(0, \phi_i^2)$.  
        \item Based on  $\bz^{*}=\{\iz_{i}^{*}: \bs_i\in \cS\}$, compute the spatial  predictions at $\cS^g$ resulting in $\widetilde{\by}$
         .
        \item Obtain the conditional simulation as   $\bv=\widehat{\by} + ({y}(\cS^g)-\widetilde{\by} )$.
    \end{enumerate}
The resulting vector $\bv$ has the correct covariance structure associated with the Kriging prediction uncertainty and multiple realizations of $\bv$ can be interpreted as all being equally plausible representations of $y({\cal S}^g)$  consistent with the observations. In more mathematical terms, $\bv$ is a draw from the conditional distribution of ${y}({\cal S}^g)$ {\it given} $\bz$ under the assumption of a Gaussian process, known covariance, and normally distributed measurement error and we identify this explicitly in  (\ref{eq:condmvn}).  It is worth mentioning that in Step 3 the nugget variance is  just $\tau^2$  ( i.e., $\tau^2 \equiv \phi_{i}^2$ ) in the exact implementation of this algorithm. However, in our approximations this variance will be inflated to reflect the additional variance of the approximations. 

Note that, the resulting vector $\bv$ is a draw from the conditional distribution assuming that all covariance parameters are known but will still have more variability than just the spatial prediction, i.e., the conditional mean.
  
\subsection{Local Kriging}\label{sec:localKriging}

\begin{figure}[h!]
    \centering
    \includegraphics[width = \textwidth]{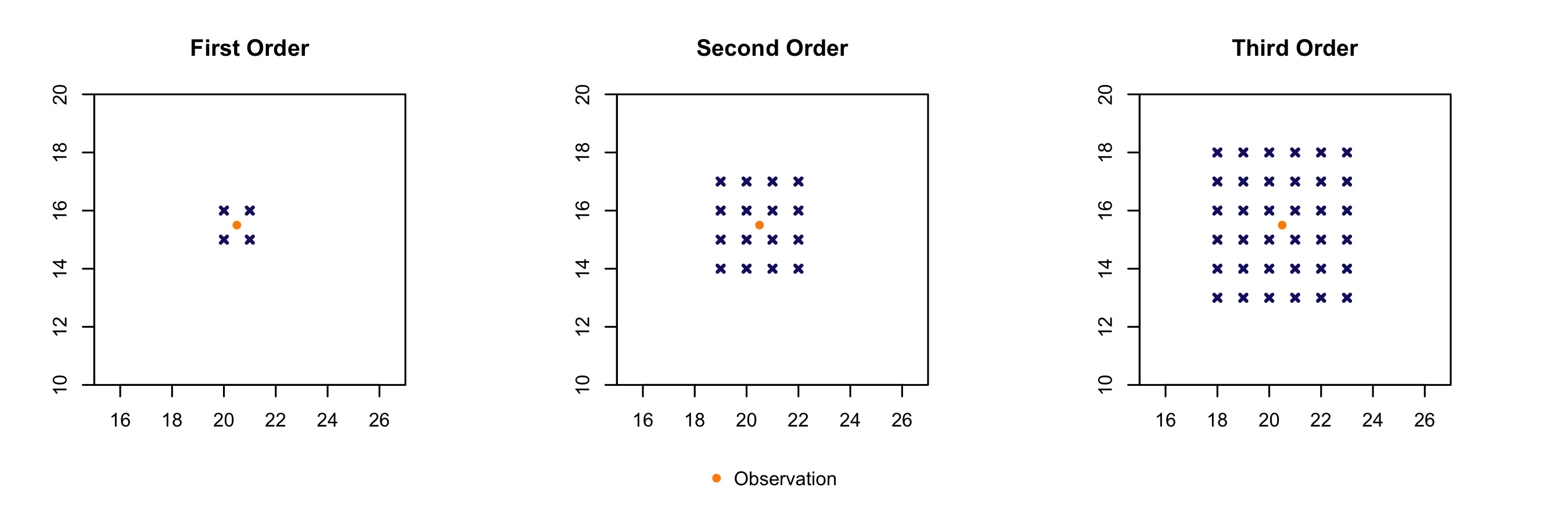}
    \caption{Illustration of the neighborhood  order $n_p =1, 2, 3$ used in the  local Kriging approximation.}
    \label{fig:2dorderneigh}
\end{figure}

The method of \textit{local Kriging} approximates Steps 3 and 4  of the conditional simulation algorithm entailing  an  unconditional  simulation  of the spatial process at the full set of locations $S=\cS^g\cup \cS$.
The approximation made by local Kriging starts with an unconditional
simulation over the grid $\cS^g$ using CE and results in the random
vector $\mathit{y}(\cS^g)$ of length $M$. These results are used to
generate off-grid observations, $\cS$. Note that, if this operation was done
using the full conditional distribution of $\mathit{y}(\cS)$ given
$\mathit{y}(\cS^g)$ then this would be exact. The difficulty, of
course,  is that it becomes computationally expensive for even
moderate size grids. Therefore, the goal is to use a limited subset
of $\mathit{y}(\cS^g)$  to generate the process at off-grid
locations.  We define a grid box as being formed from four adjacent grid points.
Denote the order neighborhood as $n_p$ where this is the
number of grid  boxes containing a particular point in each dimension. This equates to
$2n_p$ grid points closest to an off-grid observation in 1-dimension and
$(2n_p)^2$ grid points in 2-dimensions (see  Figure
\ref{fig:2dorderneigh}).  With this neighborhood scheme it is
useful to expand the range of grid points be larger than any
observation location by at least $n_p$ grid units. By adding this
margin one avoids edge effects; all neighborhoods have the same size.

Let  $\cS^g_i \subset \cS^g $ be the subset of grid points for the $n_p$ order neighborhood
surrounding  $\bs_i$  and  $\mathit{y}(\cS_i^g)$ be the subvector of 
the simulated process on the grid.  The off-grid value, $\iy(\bs_i)$ is
generated as  a draw from the conditional distribution of $\iy(\bs_i)$ given
$\mathit{y}(\cS_i^g) $.  Two important features figure into this
approximation. The first is that only the nearest neighboring grid
points are used as the conditioning set. We assert that this is not a
serious issue due to the screening effect for spatial prediction. The
second assumption is that for any pair of observations $i$ and $j$ we
assume that $\iy(\bs_i)$ and $\iy(\bs_j)$ are independent conditional on
$\mathit{y}(\cS_i^g) \cup  \mathit{y}(\cS_j^g)$. This is a strong
assumption but can also be justified  by the screening effect
combined with a fine conditioning grid and separates observation 
locations into different grid boxes.
 More discussion about this
assumption will be given Section 3.5.
Note that the assumption of conditional independence allows for
parallel simulation of the process at the observation locations given
the grid values and is different than a sequential algorithm using
Vecchia approximations to the covariance matrices.  The simulated
observations with this local Kriging method can still exhibit strong
correlations,  however,  based on the correlations among the
conditioning  grid values. 

To summarize,  the local Kriging  method proposes the following computations to approximate Steps 2,  3  and 4 in the conditional simulation algorithm.
   \begin{itemize}
       \item[(a)] Simulate the process on the regular grid,  $\cS^g$, using CE.
       \item[(b)] Using an order neighborhood of size $n_p$,
        perform univariate sampling of  $\mathit{y}(\bs_i)$  conditional on $\mathit{y}(\cS_i^g)$.
        \item[(c)] Use a nugget variance of $\phi^2_i = \tau^2 + \gamma_i$ in generating synthetic observations, where $\gamma_i$ represents the conditional variance in generating synthetic observations at off-grid locations.
        \item[(d)] Use $\tau^2$ alone for the nugget variance in the spatial prediction. 
   \end{itemize}
   
This is the algorithm studied in the next section, however, a slightly more complicated version   has some practical benefit and little extra computational overhead. If more than one observation location is in a grid box then the conditional sampling in (b) can be done  for the multivariate vector of observations in the grid box.  Some more details about this are given in Section 3.6 and can avoid working with very fine grids just to allocate all the observations into separate grid boxes. The conditional sampling for observations in {\it separate} grid boxes is still done independently. 
   
\subsection{Nearest Neighbor Kriging}\label{sec:NNk}

A second method to adapt CE for use with off-grid points will be termed \textit{nearest neighbor Kriging}. In this method, the  information in the spatial observations is shifted to the closest grid locations. In this way there are no longer any off-grid locations and CE alone can be used in Step 2 of the conditional simulation process.  The simplest way to shift the data is to identify the closest grid point and then just use that as the surrogate location. We make this operation more robust by a spatial prediction from the full set of observations to the $n_p$ order neighborhood for each observation location. This is illustrated by referring  again to Figure \ref{fig:2dorderneigh}. In addition, the  nugget variance, nominally $\tau^2$ is inflated at each grid point observation by the error variance from this prediction.  The net result is a surrogate spatial data set that attempts to retain the information from the  original, irregular locations but is now registered to locations on a regular grid.  This transformation of the data and the locations is only done once for each choice of spatial model and then the standard conditional simulation algorithm listed above can be used without modification.  
  
    A benefit to this approximation is its simplicity and in the case of very fine grids and a sparse set of observation locations one would expect this approach to be very effective.  However, in general, artificially increasing the number of observations adds to the computation and grids that are close to the resolution of the observation spacings may introduce biases.   Here the highlighted grid points  can be thought of as the new ``observations" which will be used in place of the original  data  locations and with some additional error included in the observational model.  Another disadvantage is that the redistribution of the observations to the grid points does not correspond to a specific approximation to the conditional distribution from the off-grid points \cite{cressie1993} and can result in biases. 

    \section{Evaluation of approximation of  prediction errors}\label{sec:derivation}
    
    For the purposes of this work, the exact standard error for Kriging and approximate standard errors for local and nearest neighbor Kriging have closed forms. These exact expressions avoid the additional complication of a simulation study to assess the accuracy of the approximations. They also give some insight into how the approximations are made. 
    
   The following matrices will be used to derive the  explicit formulas in this section. Throughout the subscripts 1 and 2 will refer  respectively to the grid and observation subsets. First we have the full covariances and cross-covariance matrices for the grid points and observations:
    \begin{eqnarray}
    \underset{(M \times M)}{K_{11}}&= & \operatorname{Cov}(y(\cS^g),y(\cS^g)) = C(\cS^g,\cS^g) \\
    \underset{(M \times n)}{K_{12}} &=& \COV(y(\cS^g),y(\cS) = C(\cS^g,\cS) \\
    \underset{(n \times M)}{K_{21}}&= &K_{12}^T  \nonumber  \\
     \underset{(n \times n)}{K_{22}}&=& \COV(y(\cS), y(\cS)) = C(\cS, \cS). \label{eq:covmatrices} 
     \end{eqnarray}
    Also, the conditional distribution for the multivariate normal  follows the well-known expression 
 \begin{equation}
    [ y(\cS^g) | \bz ] \sim MN(  K_{12}(K_{22}+ \tau^2 I) ^{-1} \bz , K_{11}-K_{12}(K_{22}+ \tau^2 I)^{-1}K_{21})\label{eq:condmvn}
    \end{equation}
    and the conditional simulation algorithm is an efficient way to sample from this distribution. 
 Moreover,  the diagonal elements for the conditional covariance are the appropriate prediction variances. We study whether these variances are accurately approximated by the local Kriging and nearest neighbor Kriging methods.

    \subsection{Local Kriging Approximate Conditional Variance}
    
    To derive the approximate conditional variance for local Kriging, we first identify the weights to predict the spatial process on  $\cS$ based on the values at  $\cS^g$.  In this approximation it is helpful to define an $n\times M$  incidence matrix, $\cJ$, of 0 and 1 values such that 
    $\cJ_{jk}$ is 1 for all $n_p$ order neighboring grid points of $\bs_j$ and zero otherwise.   
  It is also useful to define subsets of the full covariance matrices based on these neighborhoods. Given a specific neighborhood size, $n_p$ , let $\cS^g_j$ be the grid locations comprising the neighborhood of $\bs_j$ and we have the covariance (sub)matrices 
  \[  K_{22}^i =\operatorname{Var}( y(\bs_i)) \mbox{ ,  }  {K_{11}^i} =\operatorname{Cov}(y(\cS^g_i),y(\cS^g_i)) \mbox{, and  } {K_{21}^i} =\operatorname{Cov}(y_i, y(\cS^g_i))  \]
     We have the conditional mean and variances for the $j^{th}$ observation given by 
   \[  \widehat{y}_j = K_{21}^i( K_{11}^i)^{-1}y(\cS^g_i)= K_{21}^i( K_{11}^i )^{-1} \cJ y(\cS^g) = W_{1}^i y(\cS^g), \]
   where $W_1^i $ is a vector of weights that map the  grid values to the conditional mean for the observation. 
   Also we have the conditional variance based on the local values 
   \[  \gamma_i  = K_{22}^i - K_{21}^j( K_{11}^i )^{-1}( K_{12}^i)^T. \]
Given these individual estimates let $W_1$ denote the $n \times M$ matrix resulting from stacking $\{ W_{1}^i \}$ as row vectors and 
$\gamma$ a vector representing $\{ \gamma_i \}$. 
  With this notation the synthetic observation vector at Step 3  in the conditional algorithm is approximated as 
  \[  \bz^* = { \by^* }   + \varepsilon^*, \]
  where  ${ \by^* } = W_1  y(\cS^g)$ and  $\varepsilon^*$ are  independent and  normally distributed with variance $\phi^2_i = \gamma_i + \tau^2$.

 The {\it weights} for predicting  the grid points from these  synthetic observations  (Step 4) should only incorporate the  actual nugget variance for  the  original spatial data  and can be written as:
    \begin{equation*}
        W_2 = K_{12}^T(K_{22}+\tau I)^{-1},\label{w2full}
    \end{equation*}
    \noindent where $K_{12}$ and $K_{22}$ are defined as in Eqn.~\eqref{eq:covmatrices}. We can now define the composition of these 
    weights in the form of the matrix $\Lambda$:
    \begin{equation*}
    \underset{M\times M}{\Lambda} = W_2W_1. \label{eq:lambdafull}
    \end{equation*}
    Finally, we can define the predicted values on the grid  from Step 4 of the conditional simulation  as $\widehat{\mathit{\by}}^{S* }= \Lambda \mathit{y}(\cS^g)+W_2\varepsilon^*$ and 
    $\mathit{y}(\cS^g)$ as the ``true" process from CE. We can write:
    \begin{equation}
        \widehat{\by}^{S*} - y(\cS^g) = \Lambda y(\cS^g) -
        y(\cS^g) + W_2(\varepsilon_{W_1} + \varepsilon_{\tau}) =
        (\Lambda - I) y(\cS^g) + W_2( \varepsilon^* ), 
    \end{equation}
 and from this we derive a formula for $ \VAR( \widehat{\by}^{S*} - \by^{S^g}) $ assuming $y(\cS^g)$ and $\varepsilon^*$ are independent. Note that this computation builds in the approximation that observations in separate grid boxes are assumed to be conditionally independent given the neighboring grid points. 
  
    \subsection{Nearest Neighbor Approximate Conditional Variance}
    
    The first step of this method is to translate the off-grid observations into a surrogate problem with observations at grid locations that are nearest neighbors to the observation locations. We will assume there are $n_{\cal N}$ total number of surrogate locations. In the case of no overlapping neighborhoods  and in 2 dimensions $n_{\cal N} = n  (2 n_p)^2$, the number of observations multiplied by the neighborhood size.   Accordingly, let $\cJ_{\cal N}$ be an $n_{\cal N}  \times M$ incidence  matrix, consisting of  0 and 1 values that maps the full grid locations into just the surrogate observations on the grid.  The weight function mapping observations to this subset of the grid is given by 
    \begin{align}
        W_1^{\cal N} &= \cJ_{\cal N} K_{21}(K_{11}+\tau^2I)^{-1}. \label{eq:NNkweight1}
    \end{align}
   and the covariance matrix for the prediction errors associated with this mapping is the submatrix:
     \[  \cJ_{\cal N}   ( K_{11}-K_{12}K_{22}^{-1}K_{21} ) \cJ_{\cal N}^T, \]
  Lastly,  let $\sigma_{\cal N}^2$ be the diagonal elements of this matrix.
   The synthetic observations  in Step 3 of  the conditional simulation follow 
   \[  \bz^* =  \cJ_{\cal N} y(\cS^g)  + \varepsilon^*,  \]
   where $\varepsilon^*$   are  independent and  normally distributed with variances $\sigma_{\cal N}^2 + \tau^2 $.
   
    The uncertainty estimated by Monte Carlo from the approximate conditional simulation is  just the 
    conditional variances from the surrogate model, i.e.,  
    \[ K_{22} - K_{22} \cJ_{\cal N} { (   K_{\cal N} + \tau^2 I + \Sigma _{\cal N}) ^{-1} \cJ_{\cal N}^T  K_{22} }. \]
    and $\Sigma _{\cal N}$ the diagonal matrix with entries $\sigma_{\cal N}^2$.
    
\subsection{Analysis of the approximations to the prediction variance}

In this work covariance functions from the Mat\'ern covariance family are considered:
\[  C_{\nu}(d) = \sigma^2\frac{2^{1-\nu}}
{\Gamma(\nu)}\left( \sqrt{2\nu}\frac{d}{\theta} \right)^{\nu}K_{\nu}\left(\sqrt{2\nu}\frac{d}{\theta}\right), \] 
where $\Gamma$ is the gamma function, $K_{\nu}$ the modified Bessel function of the second  kind, $\nu$ the smoothness parameter, and $\theta$ the range or scale parameter. Two cases for $\nu$ are considered: (i) Exponential covariance with $\nu=1/2$ and (ii) Mat\'ern covariance with $\nu=3/2$. These $\nu$ values have been chosen for their closed form representation in the Mat\'ern covariance family, as the $1/2$  integer cases do not require evaluating a Bessel function. 

It is assumed that the variance of the stochastic process, also
 known as the ``sill" of the process,
 $\operatorname{Var}(y(\bs_i))=\sigma^2=1$. While the sill is
 assumed to be constant, examining various values for $\tau^2$ (or
 $\tau$) will give a sense for how the method performs according to a
 diverse set of ratios between the sill, $\sigma^2$, and the nugget
 variance, $\tau^2$. This relationship is formalized with the
 parameter $\lambda=\tau^2/\sigma^2$, called the smoothing parameter.
 Moreover, $1/\lambda$ can be recognized as the more conventional
 signal-to-noise ratio common in signal processing. We consider three nugget values $\tau^2=0.1^2, 0.2^2, 0.3^2$.
    
Three range parameters are also considered for each case of smoothness so that $\theta$ could be a short, medium or longer range parameter relative to the domain. Specifically, the range parameters are set so that for a given smoothness, the correlation between two observations falls to 5\% at a distance of 20, 45, or 70 units. These values are used for their applicability to earthquake IM ranges, as seen in \cite{loth2013spatial}. Thus the numerical design has three factors comprising $18 = 2 \times 3 \times 3 $ separate cases.  We consider a 2 dimension, square spatial domain with extent $[0,60] \times [0,60]$. 

The focus of this study is on the accuracy of the prediction standard errors and we use relative percent error as a measure of how well the approximations work against the exact formula. For this reason a  moderate grid size  is used so that the exact standard errors can be readily computed.  
 The locations of the observations are chosen  so that both the $x$ and $y$ coordinates are uniformly distributed on the interval $[0,60]$. One hundred different random configurations are used in this experiment to ensure that the relative errors analyzed are not due to chance from a single configuration. For the numerical analysis in this section, there are 35 observations on a grid size of $61 \times 61$. The ratio of number of observations to the size of the grid  reflects the density of observations  and grid for  the USGS application to ShakeMap  data. It is also    a case where a much finer prediction grid  is considered relative to the spacing of the observations.  Although this is a small number of observations these results should be consistent for larger domains with the same relative observation and grid densities. 
\subsection{Numerical Results}
\begin{figure}[htbp]
    \centering
    \begin{tabular}{cc}
        \includegraphics[height = 0.35\textheight, width=.5\textwidth]{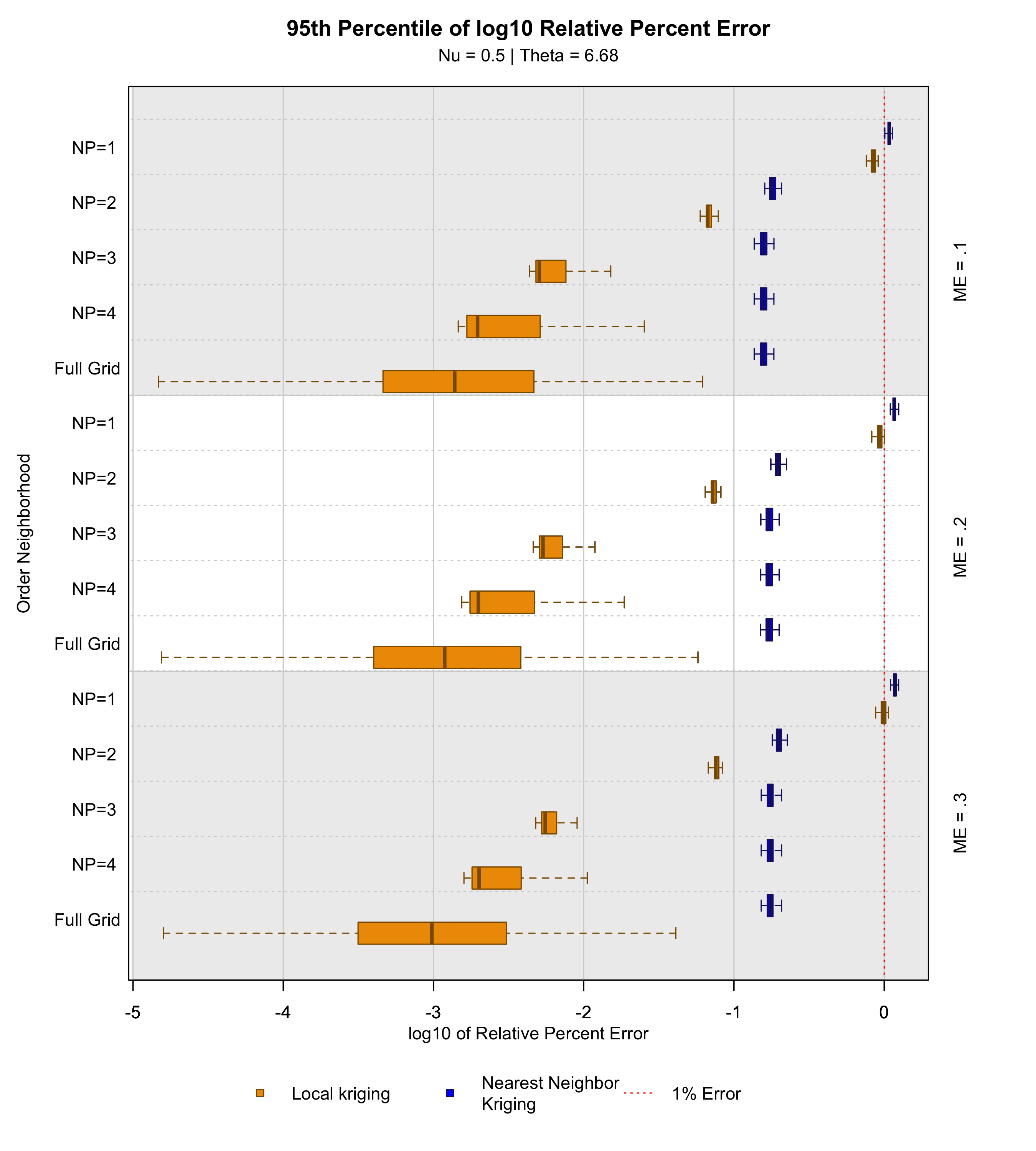}
         & \includegraphics[height = 0.35\textheight, width=.5\textwidth]{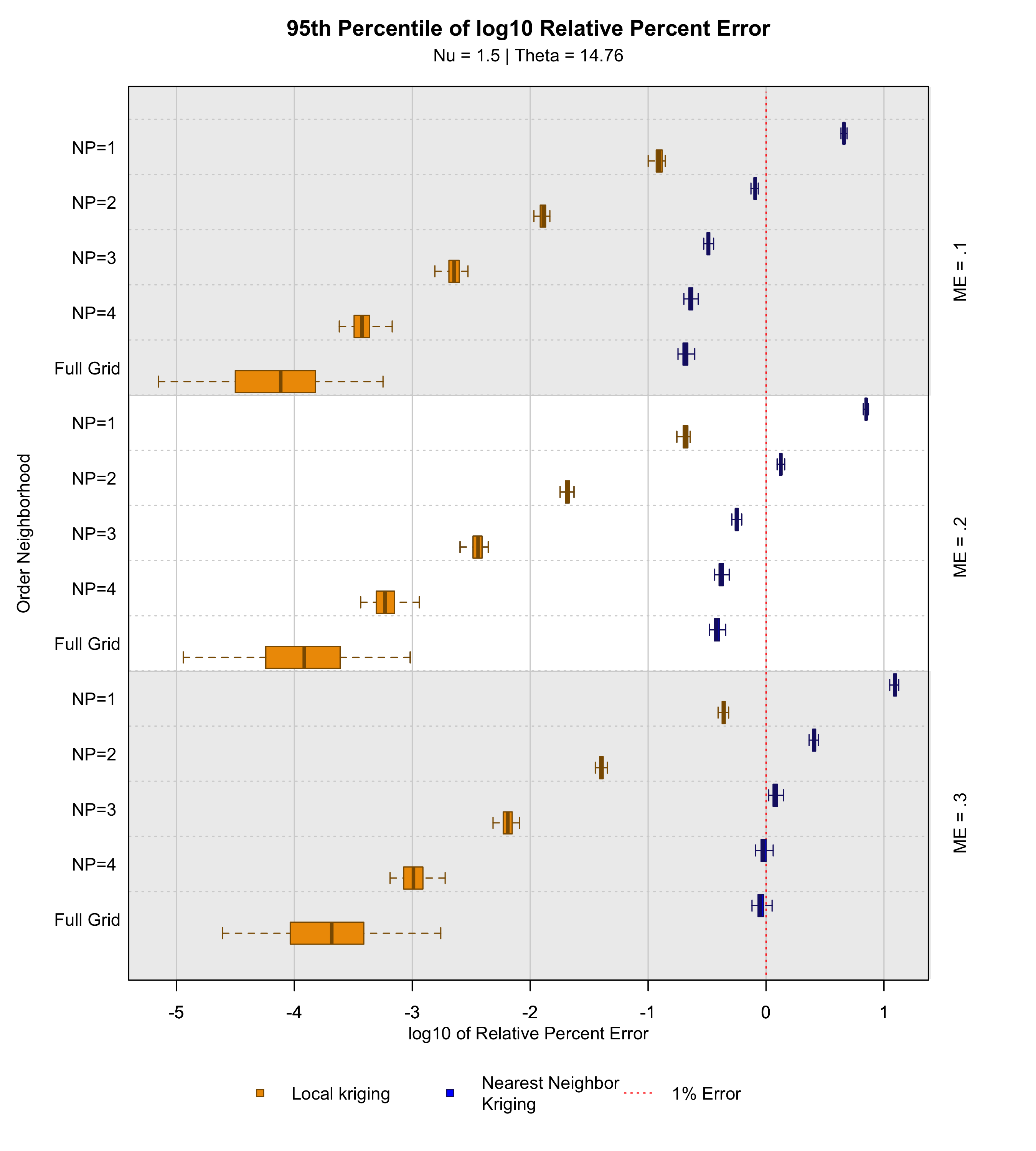} \\
         (a) & (b)
    \end{tabular}
    \caption{Relative percent difference between local Kriging, nearest neighbor Kriging and simple Kriging for $\nu=0.5$ and $\theta=6.68$.}
    \label{fig:relpctboxplot}
\end{figure}

For each combination of smoothness parameter, range parameters and configuration of observation locations the exact prediction  standard error ( $SE_E(\bs)$), the   standard error based on the local approximation ( $SE_L(\bs)$) and the   standard error based on the nearest neighbor approximation ($SE_N(\bs)$) are computed for the  $61^2 =3721$ grid locations and we find these errors for neighbor sizes $n_p = 1, 2,3,4$. 
In particular, the relative percent errors are defined  as  
$$
E_L (\bs) = 100 \times ( SE_L(\bs) - SE_E(\bs))/SE_E(\bs),\ \mbox{and},\ 
E_N(\bs) = 100 \times ( SE_N(\bs) - SE_E(\bs))/SE_E(\bs).
$$
The results are summarized in Figure~\ref{fig:relpctboxplot}, by boxplots of the  95th percentile (over the grid locations) of the  errors for  the 100 configurations of observations.  The dashed red line represents the 1\% relative error line. It is clear from this graph that the nearest neighbor method does not improve after an order neighborhood of 3 . On the other hand,  $E_L (\bs)$ appears to decrease with $n_p$ but exhibits more variability.  Regardless of these differences, for both methods the error quantiles tend to remain under 1\%.

Another way to score the closeness of the approximations to the exact prediction variances  is by quantifying their agreement to a specific number of significant digits. This comparison is motivated by  a prediction confidence interval being computed to a practical accuracy. For example the width is accurate to  three digits or about  .1\%  Table \ref{lkSigFigExp} shows the percent of  local  standard errors that are correct to three significant figures when compared to the exact prediction standard error for an exponential covariance. Table \ref{nnkSigFigExp} shows the same but for the nearest neighbor method. Note that, even when using a full grid for local Kriging, there is still 1 configuration  that is not correct to three significant figures. Table \ref{nnkSigFigExp} shows that nearest neighbor method does not do as well even when a full grid is used.

\begin{table}
    \centering
    \bgroup
    \def\arraystretch{1.25}
    \caption[Fraction of standard errors correct to 3 significant figures in 2D ($\nu=0.5$)]{Portion of standard errors resulting from local Kriging correct to 3 significant figures when compared to simple Kriging error for an exponential covariance function across 50 configurations of the grid. Values of $\theta$ are chosen so that the correlation between two observations falls to 5\% at a distance of 20, 45, or 70 units, which has applicability to earthquake IM ranges, as seen in \cite{loth2013spatial}.}

    \vspace{1em}
    \begin{tabular}{|c|c|cc|cc|cc|cc|}
    \hline
   \multirow{2}{2em}{$\theta$} & \multirow{2}{2em}{$\tau$} & \multicolumn{2}{c|}{$n_p=1$} & \multicolumn{2}{c|}{$n_p=2$} & \multicolumn{2}{c|}{$n_p=3$} & \multicolumn{2}{c|}{Full}\\\cline{3-10}
    & & Mean & Sd & Mean & Sd & Mean & Sd & Mean & Sd\\
   \hline
    \multirow{3}{2em}{6.68} & 0.1 &  0.115 & 0.034 & 0.736 & 0.020  & 0.974 & 0.009 & 0.994 & 0.009\\
         & 0.2 & 0.117 & 0.035 &  0.746 & 0.020 & 0.974 & 0.012 & 0.993 & 0.012 \\
         & 0.3 & 0.123 & 0.037 & 0.759 & 0.019 & 0.974 & 0.013 & 0.993 & 0.013 \\
    \hline
    \multirow{3}{2em}{15.02} & 0.1 & 0.066 & 0.019 & 0.741 & 0.018 & 0.970 & 0.010 & 0.994 & 0.010\\
         & 0.2 & 0.070 & 0.021  & 0.756 & 0.018 & 0.970 & 0.013 & 0.993 & 0.013 \\
         & 0.3 & 0.074 & 0.022 & 0.773 & 0.019 & 0.972 & 0.014 & 0.993 & 0.013 \\   
    \hline
    \multirow{3}{2em}{23.37} & 0.1 & 0.069 & 0.019  & 0.777 & 0.015 & 0.973 & 0.01 & 0.995 & 0.010 \\
         & 0.2 & 0.075 & 0.022  & 0.791 & 0.017 & 0.973 & 0.013 & 0.994 & 0.012 \\
         & 0.3 & 0.085 & 0.025 & 0.812 & 0.017 & 0.974 & 0.014 & 0.993 & 0 .013\\   
    \hline
    \end{tabular}
    \label{lkSigFigExp}
    }
\end{table}

\begin{table}
    \centering
    \bgroup
    \def\arraystretch{1.25}
    \caption[Portion of standard errors correct to 3 significant figures in 2D ($\nu=1.5$)]{Portion of standard errors resulting from nearest neighbor Kriging correct to 3 significant figures when compared to simple Kriging error for an exponential covariance function across 50 configurations of the grid.}
    \vspace{1em}
    \begin{tabular}{|c|c|cc|cc|cc|cc|}
    \hline
   \multirow{2}{2em}{$\theta$} & \multirow{2}{2em}{$\tau$} & \multicolumn{2}{c|}{$n_p=1$} & \multicolumn{2}{c|}{$n_p=2$} & \multicolumn{2}{c|}{$n_p=3$} & \multicolumn{2}{c|}{Full}\\\cline{3-10}
    & & Mean & Sd & Mean & Sd & Mean & Sd & Mean & Sd\\
   \hline
    \multirow{3}{2em}{6.68} & 0.1 &  0.228 & 0.036 & 0.896 & 0.003  & 0.917 & 0.003 & 0.918 & 0.003\\
         & 0.2 & 0.215 & 0.035 &  0.895 & 0.003 & 0.915 & 0.003 & 0.918 & 0.003 \\
         & 0.3 & 0.211 & 0.036 & 0.896 & 0.004 & 0.916 & 0.003 & 0.919 & 0.003 \\
    \hline
    \multirow{3}{2em}{15.02} & 0.1 & 0.006 & 0.024 & 0.840 & 0.006 & 0.901 & 0.003 & 0.904 & 0.004\\
         & 0.2 & 0.003 & 0.019  & 0.830 & 0.006 & 0.900 & 0.004 & 0.902 & 0.003 \\
         & 0.3 & 0.074 & 0.016 & 0.773 & 0.007 & 0.972 & 0.004 & 0.993 & 0.004 \\   
    \hline
    \multirow{3}{2em}{23.37} & 0.1 & 0.019 & 0.015  & 0.785 & 0.009 & 0.889 & 0.004 & 0.894 & 0.004 \\
         & 0.2 & 0.007 & 0.009  & 0.759 & 0.010 & 0.884 & 0.005 & 0.890 & 0.005 \\
         & 0.3 & 0.004 & 0.006 & 0.738 & 0.009 & 0.882 & 0.004 & 0.890 & 0.004\\   
    \hline
    \end{tabular}
    \label{nnkSigFigExp}
    }
    
\end{table}

\subsection{Model misspecification}
\label{sec:model-miss}
\begin{figure}
    \centering
    \includegraphics[height =2in, width=2.5in]{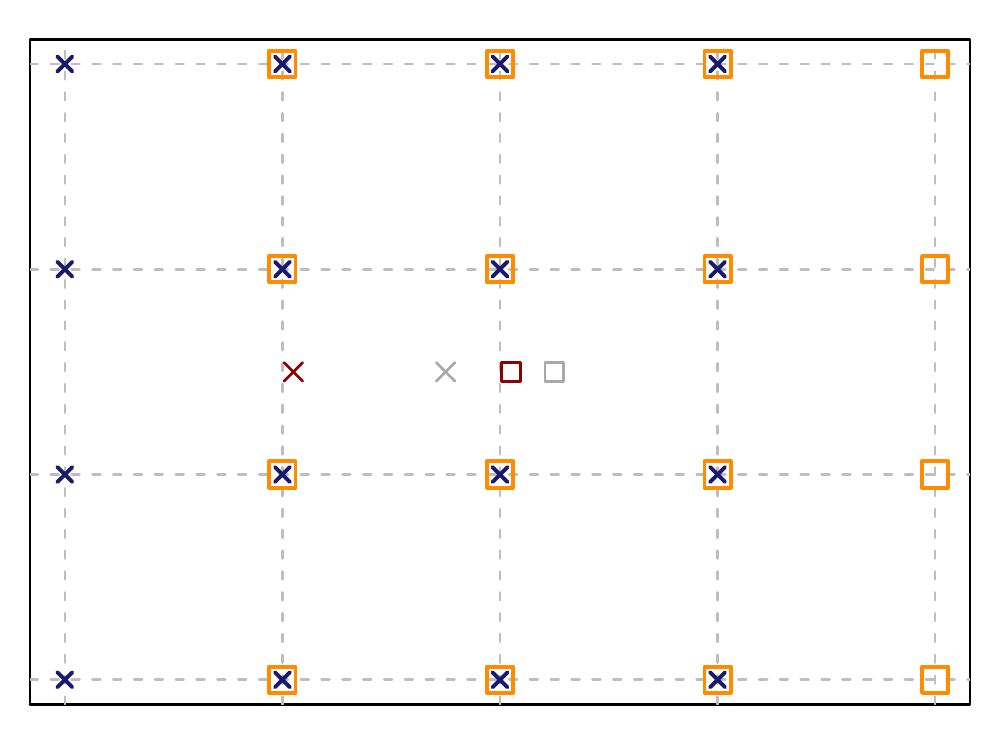}
    \caption{Location of grid and pairs of off-grid points for testing. Both sets of points have a second order neighborhood plotted where hollow orange squares correspond to the right red and grey symbols while the dark blue X's to the left red and grey symbols.}
    \label{fig:locgridpairs}
\end{figure}

A key assumption in these approximate methods is that the off-grid observations are simulated separately and the prediction errors added to the off-grid simulated values are assumed to be independent from one another. This approach is very convenient for  parallel computation and in many cases can be assumed due to the fine resolution of the simulated process on the grid by circulant embedding. In this section we give some justification for assuming independence that we believe is tied to the screening effect for fine grids that separate observations.  We consider a grid with unit spacing, second order neighbors, and  two observations that are 1/2 and  1 unit apart. We  position these observation pairs to give the largest conditional correlations, as seen in Figure~\ref{fig:locgridpairs}.  The conditional covariance is found for the conditional distribution of these two pairs for the Mat\'ern family  of covariances. We vary the range parameter in the interval  $[0.2, 10]$ and the smoothness in the interval $[0.5, 1.5]$ to obtain a maximum conditional correlation of 0.181 for one unit spacing (red circles in Figure~\ref{fig:locgridpairs}) and 0.293 for half unit spacing (grey squares in Figure~\ref{fig:locgridpairs}). These results suggest that it is important to have the grid separate observation (off-grid)  locations,  however, the screening effect greatly reduces the correlation for locations centered in adjacent grid boxes. As might be expected,   these correlations increase  somewhat as the smoothness of the Mat\'ern family is increased.

\subsection{Fast Computation}

The basic off-grid prediction involves a small linear combination of neighboring grid values for each off-grid location. This can be easily converted to a parallel operation by distributing the off-grid computation to multiple cores and tiling the grid locations to avoid passing  the entire gridded field to each core. In this study, however,  a serial  approach was taken to assess how a simpler coding of the algorithm would perform. The basic idea is to rephrase the off-grid prediction in terms of a sparse matrix multiplication of  the gridded field value and to leverage efficiency from the {\tt spam} sparse matrix package in R. 

Recall that the off-grid prediction involves a set of weights based on the nearest neighbor grid points. This can be divided up into two matrices. Recall $K_{12}^j$ is the cross covariance matrix between the nearest neighbors and the off-grid location $\bs_j$. We note that for regular grids finding nearest neighbors  and their indices is particularly fast because it is equivalent to finding the nearest integer values to real numbers.   $K_{22}^j$ be the covariance matrix for the $n_p$ order  nearest neighbors. Since we are assuming a stationary covariance function  this matrix is the same for every observation location. 
By careful indexing one can populate the rows of a {\it sparse} matrix version of $W_1$  using as  rows   $K_{12}^j  (K_{22}^j)^{-1}$ such 
that 
\begin{equation*}
    \widehat{y}(\cS) = W_1 y(\cS^g)
\end{equation*}

Here the off-grid predictions,  also the conditional mean values, are obtained by a single sparse matrix multiplication applied to the gridded field values unrolled as a vector. The advantage of this approach is that it exploits the efficient multiplication and indexing across large vectors. 

This setup makes a few important assumptions. First, the same  neighborhood  structure will be used around each off-grid point regardless of where the off-grid point is within a grid box. That is, a point that is directly in the middle of a grid box will have the same neighborhood pattern as a point that is close to the corner of the grid box. This method also assumes the grid extends several grid points beyond all observations and so padding the edge of the grid may be necessary if any observations lay too close to the edge of the domain. Finally there is the assumption of stationarity so that the covariance matrix of nearest neighbors ($K_{22}^j $) can be inverted once and applied for all off-grid predictions. Although the nominal dimensions of a sparse $W_1 $ can be large, there are only $(2n_p)^2$ nonzero entries per row and so  the multiplication is quite fast. Also note that for generating many realizations of the grid and off-grid points, $W_1$  needs to be created only once and stored.  To generate the measurement/nugget errors one can phrase this a  multiplication of the measurement standard deviations by iid standard normal random variables. In the extension where a limited number of grid boxes have a small number of observations this step can also be accomplished as sparse matrix multiplication.  In the case  the sparse matrix has diagonal elements $\phi_i$  for single observations in a grid box. For multiple observations  in a grid box there are diaogonal blocks with the Cholesky decomposition of the conditional covariance matrix.

This algorithm has been implemented in serial  in the {\tt fields} R package and has surprisingly fast performance. Our timing is done on a Macbook Pro laptop with 2.3 GHz Quad-Core Intel Core i5 using {\tt R 4.1.0}, {\tt fields 12.6}, and {\tt spam 2.7-0}. Although we expected the sparse methods to be efficient the speedup over a conventional for loop in R was surprising.  Creating $W_1$ and $VAR(\cS) $  took under 0.15 seconds for a $2048\times 2048$ grid and for 6400 off-grid observations, and $n_p=4$.  The simulation step 3 for the off-grid values  took under 0.01 seconds. 

\begin{table}[htbp]
{\footnotesize
\centering
\begin{tabular}{|cccccccc|}
  \hline
m & n & CESetup & OffSetup & CE & OffGrid & predict & total \\ 
  \hline
128 & 400 & 0.02 & 0.01 & 0.02 & 0.00 & 0.24 & 0.61 \\ 
  128 & 1600 & 0.01 & 0.03 & 0.02 & 0.00 & 0.93 & 1.46 \\ 
  128 & 6400 & 0.02 & 0.14 & 0.02 & 0.00 & 3.67 & 4.79 \\ 
  256 & 400 & 0.07 & 0.01 & 0.07 & 0.00 & 0.95 & 1.53 \\ 
  256 & 1600 & 0.07 & 0.03 & 0.07 & 0.00 & 3.70 & 4.86 \\ 
  256 & 6400 & 0.07 & 0.12 & 0.07 & 0.00 & 14.52 & 17.97 \\ 
  512 & 400 & 0.39 & 0.01 & 0.53 & 0.00 & 3.73 & 5.40 \\ 
  512 & 1600 & 0.49 & 0.06 & 0.40 & 0.00 & 14.76 & 18.67 \\ 
  512 & 6400 & 0.44 & 0.12 & 0.48 & 0.01 & 58.65 & 71.13 \\ 
   \hline
\end{tabular}
\caption{Timing results for the complete conditional simulation algorithm.}
\label{table:sparsepredtiming}
}
\end{table}

To set these results in context it is also useful to time our proposed conditional simulation algorithm.  As above we use a serial implementation based on easy to use and high level functions from the \texttt{fields} package and we break up the timing into several key parts. The results are reported in Table~\ref{table:sparsepredtiming} for $n_p=4$. Here $M= m^2$  is the total grid size and $n$ is the number of off-grid locations. The off-grid locations are drawn from a uniform distribution on the spatial domain. 
For these computations we used an exponential covariance where the evaluation of the covariance matrix has been optimized somewhat using a  lower level C subroutine called from R. Times are in seconds with {\tt CESetup} being the time to setup computations for the circulant embedding, {\tt OffSetup} setup time for the off-grid predictions, {\tt CE} time for simulating a single realization on the grid, {\tt OffGrid} time to simulation a realization of the off-grid values, {\tt predict} the time to predict from the off-grid observations onto the regular grid, and finally {\tt draw} the total time to generate a single draw this algorithm. Thus generating 100 
members from the conditional distribution on a $128 \times 128$ grid and for 400 observations will take on the order of about a minute ($100 \times 0.61= 61$ seconds). We did not time the exact computation because this would involve a Cholesky decompostion of the full covariance matrix of size $16784 \times 16784$.  For the largest case the time for a 100 member ensemble will be substantial, amounting to about 2 hours. Note, however, that the time is dominated by how long it takes to compute the spatial process predictions from $n$ locations to the $m \times m$ grid. The time in generating the unconditional field, the subject of this work, is negligible compared to the prediction step.  

\section{Application to Ridgecrest, CA Earthquake}

 Earthquakes are one of the most devastating natural disasters and are the motivation for this work. After any significant earthquake, economic and fatality loss estimates are extremely important for local government officials to coordinate aid. Pinpointing the local variation in damage and loss in a timely manner is important. Therefore, to help prepare local governments, first responders, or utility companies, near real-time estimates for the intensity of shaking at specific locations are essential.  Most of the seismic loss problems typically depend on the regional distribution of intensity measures (IM), rather than intensity only at a single site.
 
 Quantifying ground-motion over a spatially-distributed region requires information on the correlation between the ground-motion intensities at different sites during a single event \cite{loth2013spatial}. To this end, several correlation models have been developed that describe the decay in correlation from intensities at one site to another with increasing separation distance. However, for a large scale application of  ShakeMap, computing the spatially correlated field using classical spatial methods is generally memory intensive and computationally expensive. Therefore, it is necessary to improve the speed and accuracy of simulating IMs across an area affected by an earthquake.
 
\begin{figure}
\centering
    \includegraphics[width=\textwidth,height=.35\textheight]{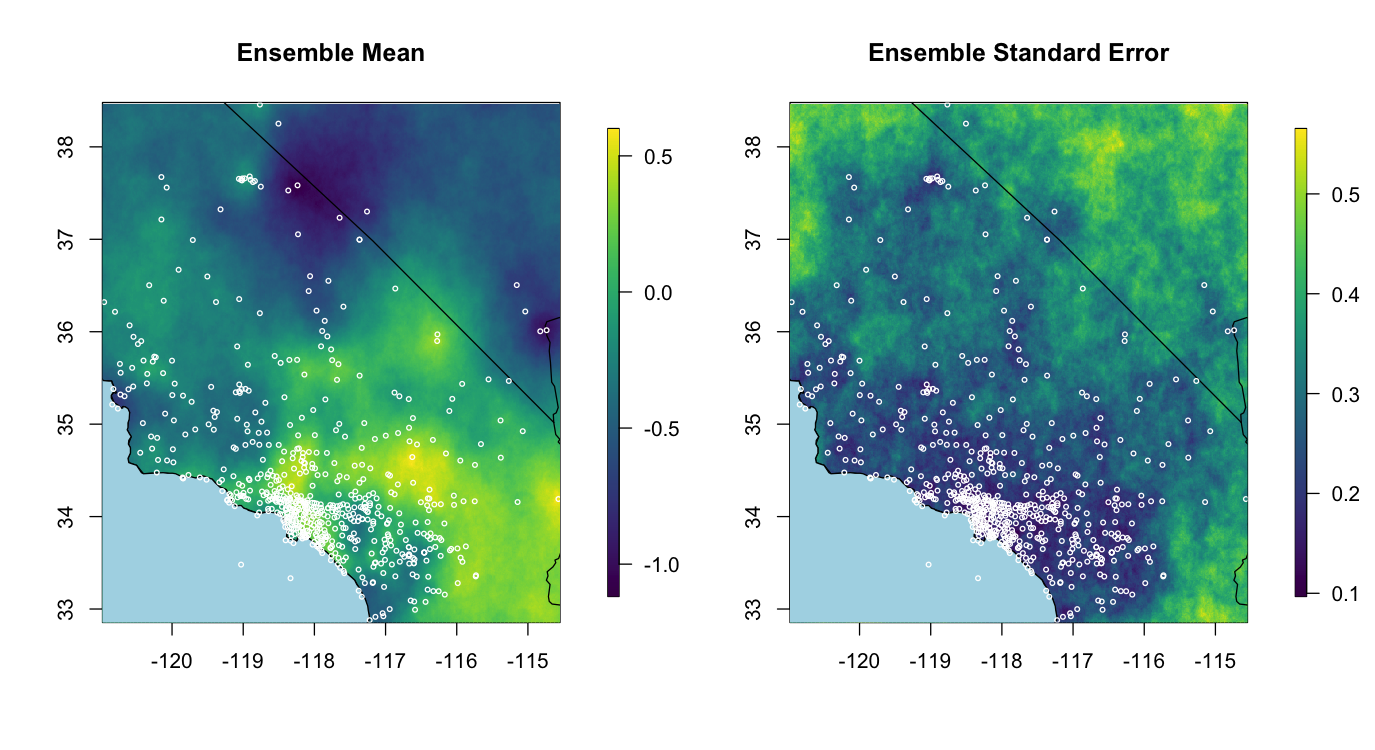}
   \caption{The mean field (left) across an ensemble of 50 for the Ridgecrest, CA earthquake using an order neighborhood of 3. The standard error (right) shows lower error around observed values and increased error as distance from an observation increases.}\label{fig:ridgecrest}
\end{figure}

To demonstrate local Kriging with observed shaking data, the algorithm is implemented using IMs observed during the 2019 Ridgecrest, CA earthquake. Here, the Ridgecrest earthquake is considered for the higher availability of station data. Observed peak ground acceleration (PGA) are normalized based on the ShakeMap Manual \cite{worden2018spatial, worden2020}. In total, 637 observed values for PGA from 633 stations are used. For stations where there are multiple recordings of PGA, the mean across the observed values is taken resulting in 633 total observed values for the event. In order for the local Kriging to be fully implemented in this example, several updates to the simulation grid had to be made. The resolution of the grid was increased 4-fold and any stations that still remained in the same box after increasing the resolution were averaged over the same grid box resulting in a single observation per grid box. The final observation count was 623 and the final prediction grid has 275,394 locations. The observations are distributed throughout southern California as seen by the white circles in Figure~\ref{fig:ridgecrest}. The left map in Figure~\ref{fig:ridgecrest} shows the final mean estimate across an ensemble of 50 realizations of the field and the right map shows the estimated standard error of the realizations. The standard error is lowest near the observation values and increases as the distance to observations increases, as expected. 

Timing of this algorithm was done on a MacBook Pro with 2.6 GHz 6-Core Intel Core i7 with 16GB of memory. Because of the increased resolution of the simulation grid, setting up the circulant matrix for simulation took a majority of the simulation time - approximately 42 seconds. However, within each run of the simulation, the unconditional simulation step took a median value of 28.6 seconds while the off-grid prediction step only took .03 seconds. Finally, the prediction to the grid took approximately 6 seconds. Thus it is clear that for larger grids where an increased resolution for the simulation grid is needed, the time for the circulant embedding setup may increase substantially. However, the off-grid prediction step remains  an extremely efficient calculation  due to the use of  sparse linear algebra.

\section{Discussion}

We have established an approximate method for conditional simulation based on local Kriging that is accurate for data analysis and also fast enough to fit into an interactive data analysis session.  A serial implementation is straightforward and easily incorporated into existing R packages for spatial data analysis. 
Based on the numerical results and timing , it is recommended that a fourth order neighborhood be used for local Kriging for both an exponential and Mat\'ern with $\nu=1.5$. Additionally, while both methods provide relative errors compared to exact Kriging that are within 1\%, it is recommended that the local Kriging method  be used. 
Sparse matrix methods provide significant speedup for the local Kriging method and it is suggested that this be used over previous methods for generating ensemble members. A significant portion of time for the entire conditional simulation algorithm is dedicated to grid prediction. However, the methods developed here address reducing time for the unconditional simulation and prediction to the off-grid points, which was successful. Future work could focus on implementing fast grid-prediction methods to further improve the conditional simulation. For example covariance functions such as the Wendland family have compact support that can be exploited for faster prediction. Some other approaches are to use fixed rank Kriging type predictions or develop approximations to the large covariance matrices needed for prediction using hierarchical matrix decompositions. 

Future work could be to implement this method in 3 dimensions for use in atmospheric or subsurface modeling. A second area of further research is to extend this analysis to anisotropic covariance function as CE can be used for anisotropic simulation.  In particular anisotropic processes could improve estimation where earthquake wave propagation is not uniform.

%% file: References/References.tex
\interlinepenalty=1000
\bibliographystyle{abbrv}
\nocite{*}
\bibliography{References/References}